\documentclass[prd,aps,preprint,amsmath,nofootinbib,amssymb,eqsecnum,showkeys,tightenlines,superscriptaddress]{revtex4-1}
\usepackage[colorlinks=true,urlcolor=blue,anchorcolor=blue,citecolor=blue,filecolor=blue,linkcolor=blue,menucolor=blue,pagecolor=blue,linktocpage=true]{hyperref} 

\usepackage[inline]{enumitem}
\usepackage[multidot]{grffile}  
\usepackage{dcolumn}
\usepackage{bm}
\usepackage{amsmath}
\usepackage{amsfonts}
\usepackage{amssymb}
\usepackage{color}
\usepackage{latexsym}
\usepackage{slashed} 
\usepackage{pstricks}
\usepackage{indentfirst}
\usepackage{mathrsfs}
\usepackage{multirow}
\usepackage{epsfig,psfrag}
\usepackage{subfigure}
\usepackage{mathtools}
\usepackage{setspace} 
\usepackage[utf8]{inputenc} 
\usepackage[scientific-notation=true]{siunitx} 

\makeatother

\allowdisplaybreaks

\begin{document}
\title{$Z$ boson mixing and the mass of the $W$ boson  }
\author{Yu-Pan Zeng}\email[]{zengyp8@mail2.sysu.edu.cn}
\affiliation{School of Electronics and Information Engineering, Guangdong Ocean University, Zhanjiang 524088, China}
\affiliation{School of Physics, Sun Yat-sen University, Guangzhou 510275, China}

\author{Chengfeng Cai}\email[]{caichf3@mail.sysu.edu.cn}
\affiliation{School of Physics, Sun Yat-sen University, Guangzhou 510275, China}

\author{Yu-Hang Su}\email[]{suyh5@mail2.sysu.edu.cn}
\affiliation{School of Physics, Sun Yat-sen University, Guangzhou 510275, China}

\author{Hong-Hao Zhang}\email[]{zhh98@mail.sysu.edu.cn}
\affiliation{School of Physics, Sun Yat-sen University, Guangzhou 510275, China}

    \begin{abstract}
	We explore the possibility of explaining the $W$ boson mass with an extra gauge boson mixing with the $Z$ boson at tree level. Extra boson mixing with the $Z$ boson will change the expression of the $Z$ boson mass, thus altering the $W$ boson mass. We explore two models in this work. We find that in the Derivative Portal Dark Matter model, there are parameters space which can give the observed $W$ boson mass, as well as the observed Dark Matter relic density. These parameters space can also fulfill the constraints from the electroweak oblique parameters and Dark Matter indirect detection. In the U(1) extension model, the kinetic mixing between extra boson and $B$ boson can also give the observed $W$ boson mass. However, to fulfill electroweak oblique parameters fit the kinetic mixing in the U(1) model can only contribute about $27~\mathrm{MeV}$ extra mass to the Standard Model $W$ boson mass. Both models indicate the extra vector boson with the best fit mass around $120~\mathrm{GeV}$.  
    \end{abstract}
\maketitle
\clearpage
\section{\label{sec:intro}Introduction}
Recently the Collider Detector at Fermilab (CDF) Collaboration has measured the mass of the $W$ boson to be $80.4335\pm 0.0094~\mathrm{GeV}$~\cite{CDF:2022hxs}, which is deviated from Standard Model (SM) prediction of $80.357\pm 0.006~\mathrm{GeV}$~\cite{ParticleDataGroup:2020ssz} and which seems to indicate new physics beyond SM. There are lots of works appeared to discuss this topic~\cite{Cirigliano:2022qdm,Borah:2022obi,Chowdhury:2022moc,Arcadi:2022dmt,Zhang:2022nnh,Mondal:2022xdy,Nagao:2022oin,Kanemura:2022ahw,Kawamura:2022uft,Peli:2022ybi,Ghoshal:2022vzo,Perez:2022uil,Zheng:2022irz,Ahn:2022xeq,Heo:2022dey,Crivellin:2022fdf,Endo:2022kiw,Du:2022brr,Cheung:2022zsb,DiLuzio:2022ziu,Balkin:2022glu,Biekotter:2022abc,Krasnikov:2022xsi,Paul:2022dds,Babu:2022pdn,DiLuzio:2022xns,Bagnaschi:2022whn,Heckman:2022the,Lee:2022nqz,Cheng:2022jyi,Bahl:2022xzi,Song:2022xts,Asadi:2022xiy,Athron:2022isz,Sakurai:2022hwh,Fan:2022yly,Zhu:2022scj,Arias-Aragon:2022ats,Cacciapaglia:2022xih,Blennow:2022yfm,Lu:2022bgw,Strumia:2022qkt,Athron:2022qpo,Yang:2022gvz,deBlas:2022hdk,Tang:2022pxh,Du:2022pbp,Campagnari:2022vzx,Zhu:2022tpr,Fan:2022dck,Han:2022juu,Alguero:2022est,Ghorbani:2022vtv,Yuan:2022cpw,Bhaskar:2022vgk}.
In this work we will explore physics beyond SM which can give the observed mass of the $W$  boson at tree level. 

In the SM, the mass of the $W$ boson and the $Z$ boson are given by the Higgs mechanism. Since the $Z$ boson is combination of the $B$ boson and the $W^{3}$ boson, which is a component of the gauge triplet $W^{i}$, the mass of the $W$ boson and the $Z$ boson are connected. Therefore it is difficult to solely change the mass of the $W$ boson. One way to alter the mass of the $W$ boson is to mix the $Z$ boson with an extra vector boson. Mix the $Z$ boson with another boson will inevitably alter the mass expression of the $Z$ boson which may alter the value of the $\mathrm{SU}(2)_L$ gauge coupling and thus the mass of the $W$ boson. There are usually two kinds of mixing: direct mixing in mass matrix and kinetic mixing. Though the normalization of the kinetic mixing terms will result in mass mixing, we will consider two models in this work: the Derivative Portal Dark Matter (DPDM) model~\cite{Zeng:2022llh} and the U(1) model~\cite{Holdom:1990xp,Lao:2020inc}. In these two models, the extra gauge boson are connected to the SM through kinetic mixing to the $Z$ boson and the $B$ boson respectively. The kinetic mixing will alter the mass expression of the $Z$ boson and thus the mass of the $W$ boson at tree level. Since electroweak oblique parameters have a strong constraint on electroweak physics, we will consider the electroweak oblique parameters constraint on these models. For the DPDM model, we also consider constraints from the observed Dark Matter (DM) relic density and DM indirect detection. 

This work is structured as follows: In Sec.~\ref{sec:general} we generally discuss the mechanism whereby the mixing between an extra boson and the $Z$ boson changes the mass of the $W$ boson. In Sec.~\ref{sec:bsm} we explore two models and discuss their capability of altering the $W$ mass. We also explore constraints from electroweak oblique parameters, DM relic density, and DM indirect detection. We conclude in Sec.~\ref{sec:con}   
\section{general discussion of prediction of the mass of the $W$  boson}%
\begin{figure}[ht]
    \centering
    \includegraphics[width=0.8\textwidth]{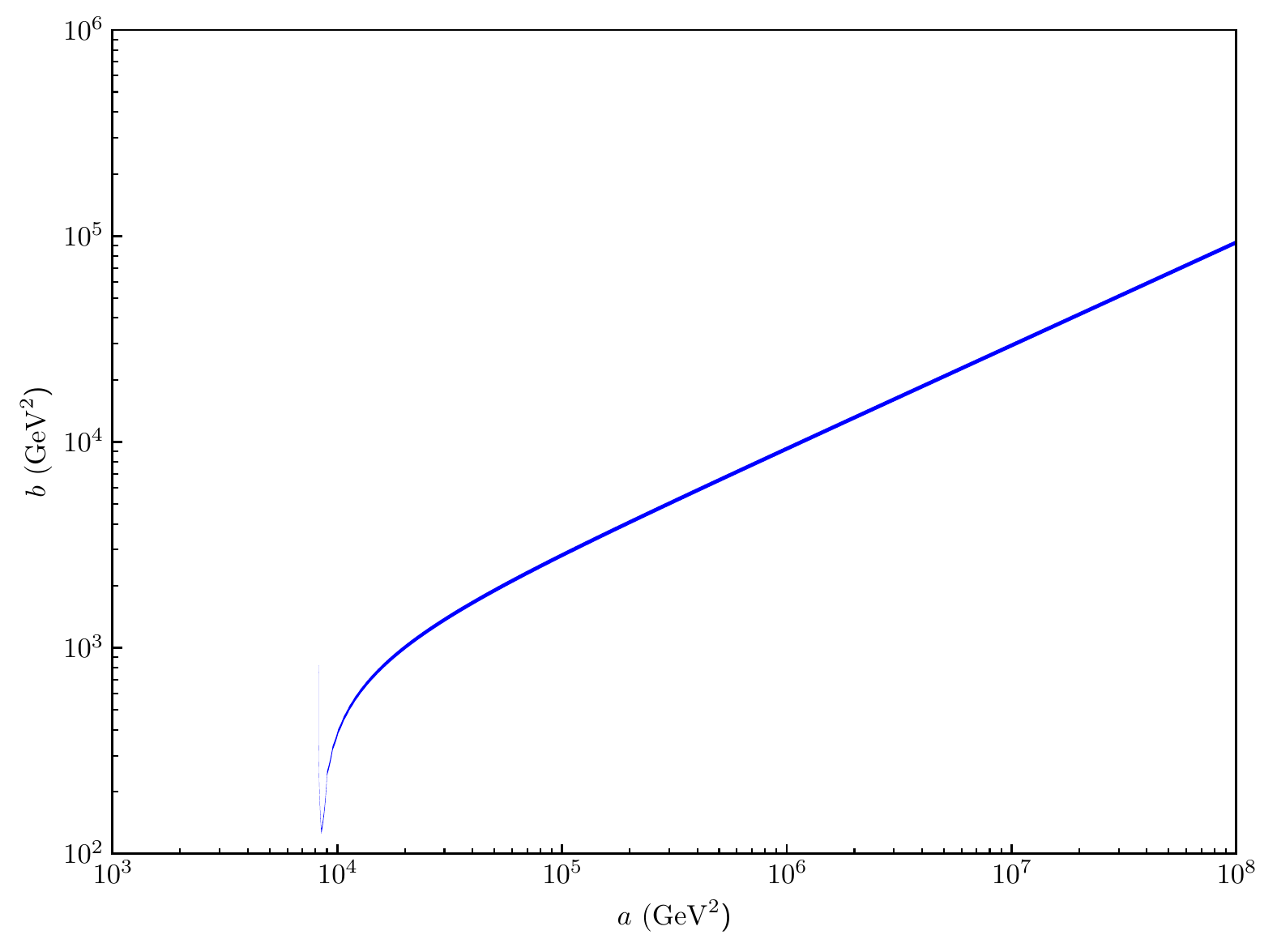}
    \caption{Band which gives the $W$ mass between $80.4053$ and $80.4617$.  }
    \label{fig:abband}
\end{figure}

\label{sec:general}
In this section we will discuss in general how an extra boson mixing with the $Z$ boson changes the mass of the $W$ boson. To see this we first write down the mass of the $W$ boson $m_{W}$  and the mass of the $Z$ boson $m_{Z}$  given by SM:
\begin{eqnarray}
    m_{W}^2=\frac{1}{4}g^2v^2,\ m_{Z}^2=\frac{1}{4}(g^2+g^{\prime 2})v^2,
\end{eqnarray}
where $g$ and $g^{\prime}$ are the gauge couplings of $\mathrm{SU}(2)_\mathrm{L}$ and $\mathrm{U}(1)_\mathrm{Y}$ gauge symmetry. And $v$ is the vacuum expectation value (vev) of the Higgs boson. When choosing the Fermi coupling constant $G_{F}$, the $Z$ boson mass $m_{Z}$ and the fine-structure constant $\alpha$ as input parameters, the $W$ boson mass will then be determined. Because the parameters involved in the $W$ boson mass can be determined by adding the following equations:  
\begin{eqnarray}
    G_{F}=\frac{1}{\sqrt{2}v^2 },\ e=\sqrt{4\pi\alpha} =\frac{gg^{\prime}}{\sqrt{g^2+g^{\prime 2}} }.
\end{eqnarray}
Going beyond the SM, we will mix the $Z$ boson with another vector boson. After that the real mass of the $Z$ will be the square root of one of the eigenvalues of the following mass matrix:
\begin{eqnarray}
    \begin{pmatrix}
	\frac{1}{4}(g^2+g^{\prime 2})v^2&b\\
	b&a
    \end{pmatrix},
\end{eqnarray}
where we have used $a$ and $b$ to denote some general mass terms. The eigenvalues of the mass matrix can be written as: 
\begin{eqnarray}
    m_{Z,Z^{\prime}}^2=\frac{1}{2}\left( \frac{1}{4}(g^2+g^{\prime 2})v^2+a\pm \sqrt{\left( \frac{1}{4}(g^2+g^{\prime 2})v^2+a \right)^2-a(g^2+g^{\prime 2})v^2+4b^2}  \right) \label{mass}.
\end{eqnarray}
Define $c=\frac{1}{4}(g^2+g^{\prime 2})v^2$, then we will get compact form of $m_{Z,Z^{\prime}}^2=\frac{1}{2}\left( a+c\pm\sqrt{(a-c)^2+4b^2}  \right)$. 
We can see the heavier mass of $m_{Z,Z^{\prime}}$ will be bigger than both $a$ and $c$, and the lighter mass of $m_{Z,Z^{\prime}}$ will be smaller than both $a$ and $c$. Therefore in order to have a bigger $c$, since observation of the $W$ mass indicates larger $g$, the value of $a$ must be lager than $c$. Therefore the mass of the $Z$ boson should correspond to the minus sign in Eq.~\eqref{mass}.  
Adopting the input parameters as $G_{F}=1.1663787\times 10^{-5}~\mathrm{GeV}^{-2},\ m_{Z}=91.1876~\mathrm{GeV},\ \alpha\approx 1/128$~\cite{ParticleDataGroup:2020ssz}, we can draw a blue band which saturates the observed mass of the $W$ boson in $3\sigma$ confidence level in Fig.~\ref{fig:abband}.

Actually we can calculate the analytic relation between $a$ and $b$ by taking the mass of $W$ boson $m_{W}$ as an input parameter. From Eq.~\eqref{mass} we can write:
\begin{eqnarray}
    b^2&&=c(a-m_{Z}^2)+m_{Z}^{4}-m_{Z}^2a\nonumber\\
	&&=\frac{4m_{W}^4}{4m_{W}^2-e^2v^2}(a-m_{Z}^2)+m_{Z}^{4}-m_{Z}^2a\label{abconstranit}.
\end{eqnarray}
Then we can give constraint on models beyond SM according to Eq.~\eqref{abconstranit}. Actually the above discussion does not take loop corrections from SM into consideration. Considering the loop corrections from SM we should replace $m_{W}$ in Eq.~\eqref{abconstranit} with $m_{W}-\delta m_{W} $, where $\delta m_{W}$ represents the loop corrections to $m_{W}$ from SM.   
\section{models beyond SM}%
\label{sec:bsm}
In this section we will explore two models beyond SM which mix the $Z$ boson with an extra vector boson and might give the observed $W$ boson mass. We also consider other constraints like electroweak oblique parameters constraint, DM relic density constraint, and DM indirect detection constraint. 
\subsection{Derivative Portal Dark Matter}
\label{sub:u_1_model}
The DPDM model extends the SM with an extra vector boson which links the dark sector and the SM through its kinetic mixing with the $Z$ boson. The relevant Lagrangian of the DPDM model can be written as~\cite{Zeng:2022llh}:
\begin{eqnarray}
    \mathcal{L}=&&-\frac{1}{4}Z^{\mu\nu}Z_{\mu\nu}-\frac{1}{4}Z^{\prime\mu\nu}Z^{\prime}_{\mu\nu}-\frac{\epsilon}{2} Z^{\mu\nu}Z_{\mu\nu}^{\prime}\\
	&&+\sum\limits_{f}  Z_{\mu}\bar{f}\gamma^{\mu}(g_{V}-g_{A}\gamma^{5})f+g_{\chi}Z_{\mu}^{\prime}\bar{\chi}\gamma^{\mu}\chi\nonumber\\
	&&+\frac{1}{2}m_{Z}^2Z_{\mu}Z^{\mu}+\frac{1}{2}m_{Z^{\prime}}^2Z_{\mu}^{\prime}Z^{\prime\mu}-m_{\chi}\bar{\chi}\chi\nonumber.
\end{eqnarray}
After normalization of the kinetic terms, the kinetic mixing between $Z$ and $Z^{\prime}$ actually result in mass mixing between them. The kinetic terms of the Lagrangian can be normalized by:
\begin{eqnarray}
   K=\begin{pmatrix}
       -k_1&k_2\\
       k_1&k_2
   \end{pmatrix},
\end{eqnarray}
where $k_1=1/\sqrt{2-2\epsilon} $ and $k_2=1/\sqrt{2+2\epsilon} $. The normalization will result in the following mass matrix between the two vector bosons: 
\begin{eqnarray}
	    \begin{pmatrix}
	    k_1^2M_1&k_1k_2M_2\\
	    k_1k_2M_2&k_2^2M_1
	    \end{pmatrix},
\end{eqnarray}
where $M_1=m_{Z}^2+m_{Z^{\prime}}^2$ and $M_2=m_{Z^{\prime}}^2-m_{Z}^2$. 
One can use an orthogonal matrix $O$ to diagonalize the mass matrix, and $O$ can be defined as 
\begin{eqnarray}
	    O=\begin{pmatrix}
		\cos \theta&\sin \theta\\
		-\sin \theta& \cos \theta
	    \end{pmatrix},\ \text{with} \tan 2\theta=\frac{2k_1k_2M_2}{(k_2^2-k_1^2)M_1}.
\end{eqnarray}
Therefore according to Eq.~\eqref{abconstranit} we can give constraint on $m_{Z^{\prime}}$ and $\epsilon$ as:
\begin{eqnarray}
    k_1k_2M_2=\sqrt{(k_1^2M_1-m_{\hat{Z}}^{2})(k_2^2M_1-m_{\hat{Z}}^{2})}\label{DPDMcons},  
\end{eqnarray}
where we have used $m_{\hat{Z}}$ to represent the experiment observed mass of $Z$ boson, which is meant to distinguish from $m_{Z}$. Also we can use the measured mass of $Z^{\prime}$ boson $m_{\hat{Z}^{\prime}}$ to reformulate Eq.~\eqref{DPDMcons}:
\begin{eqnarray}
    m_{Z}^2=\frac{1}{8k_1^2k_2^2}(m_{\hat{Z}}^2+m_{\hat{Z}^{\prime}}^2)-\sqrt{\frac{1}{64k_1^{4}k_2^{4}}(m_{\hat{Z}}^2+m_{\hat{Z}^{\prime}}^2)^2-\frac{1}{4k_1^2k_2^2}m_{\hat{Z}}^2m_{\hat{Z}^{\prime}}^2}.  
\end{eqnarray}

Apart from giving mass to the $W$ boson, we will also calculate the tree level $S,T,U$ constraints on this model. The neutral-current coupling between the $Z$ boson and SM fermions in the DPDM model can be written as:
\begin{eqnarray}
    L_{NC,\hat{Z}ff}&&=\sum\limits_{f}  (-k_2\sin \theta -k_1\cos \theta) \hat{Z}_{\mu}\bar{f}\gamma^{\mu}(g_{V}-g_{A}\gamma^{5})f\\
	      &&=\sum\limits_{f}  (-k_2\sin \theta -k_1\cos \theta) \hat{Z}_{\mu}\bar{f}\gamma^{\mu}\frac{e}{s_{w}c_{w} }(T^{3}_{f}\frac{1-\gamma^{5}}{2}-Q_{f}s_{w}^2 )f,
\end{eqnarray}
where $\hat{Z}_{\mu} $ is the mass eigenstate of the $Z$ boson. We see the above coupling is the same as that in the SM except for an extra factor $(-k_2\sin \theta -k_1\cos \theta)$, and the form of the charged-current in the DPDM is the same as that in the SM. 
Using the effective-lagrangian techniques given by ~\cite{burgess1994model}:
\begin{small}
\begin{eqnarray}
    &&\mathcal{L}_{CC, Wff}=-\frac{e}{\sqrt{2} \hat{s}_{w}}(1-\frac{\alpha S}{4(\hat{c}_{w}^2-\hat{s}_{w}^2)}+\frac{\hat{c}_{w}^2\alpha T}{2(\hat{c}_{w}^2-\hat{s}_{w}^2)}+\frac{\alpha U}{8\hat{s}_{w}^2})\sum\limits_{ij}V_{ij}\bar{f}_{i}\gamma^{\mu}\gamma_{L}f_{j}W_{\mu}^{\dagger}+\mathrm{c.c.}\\  
&&\mathcal{L}_{NC, \hat{Z}ff}=\frac{e}{\hat{s}_{w}\hat{c}_{w}}(1+\frac{\alpha T}{2})\sum\limits_{f}\bar{f}\gamma^{\mu}[T^{3}_{f}\frac{1-\gamma^{5}}{2}-Q_{f}(\hat{s}_{w}^2+\frac{\alpha S}{4(\hat{c}_{w}^2-\hat{s}_{w}^2)}-\frac{\hat{c}_{w}^2\hat{s}_{w}^2\alpha T}{\hat{c}_{w}^2-\hat{s}_{w}^2})]f\hat{Z}_{\mu},
\end{eqnarray}
\end{small}
where $\hat{s}_{w}=\sin \hat{\theta}_{w}$ and $\hat{c}_{w}=\cos \hat{\theta}_{w}$ and they are defined by:   
\begin{eqnarray}
    \hat{s}_{w}\hat{c}_{w}m_{\hat{Z}}=s_{w}c_{w}\frac{1}{2}\sqrt{ g^2+g^{\prime 2}}v=\frac{1}{2}ev=s_{w}c_{w}m_{Z}.\label{swcwrel}
\end{eqnarray}
Now we can write $S,\ T$ and $U$ in the DPDM model as
\begin{eqnarray}
    \alpha T=2(\frac{\hat{s}_{w}\hat{c}_{w}}{s_{w}c_{w}}(-k_2\sin \theta-k_1\cos \theta)-1)\\
    \alpha S=4\hat{c}_{w}^2\hat{s}_{w}^2\alpha T+4(\hat{c}_{w}^2-\hat{s}_{w}^2)(s_{w}^2-\hat{s}_{w}^2)\\ 
    \alpha U=8\hat{s}_{w}^2(\frac{\hat{s}_{w}}{s_{w}}-1+\frac{\alpha S}{4(\hat{c}_{w}^2-\hat{s}_{w}^2)}-\frac{\hat{c}_{w}^2\alpha T}{2(\hat{c}_{w}^2-\hat{s}_{w}^2)}).
\end{eqnarray}
Then we constrain the DPDM model with global fit results given by table five of Ref.~\cite{deBlas:2022hdk}:
\begin{eqnarray}
    S=0.005\pm 0.097,\ T=0.04\pm 0.12,\ U=0.134\pm 0.087,
\end{eqnarray}
with the correlation coefficient $\rho_{ST}=0.91,\ \rho_{SU}=-0.65,\ \rho_{TU}=-0.88$.  

\begin{figure}[ht]
    \centering
    \includegraphics[width=0.8\textwidth]{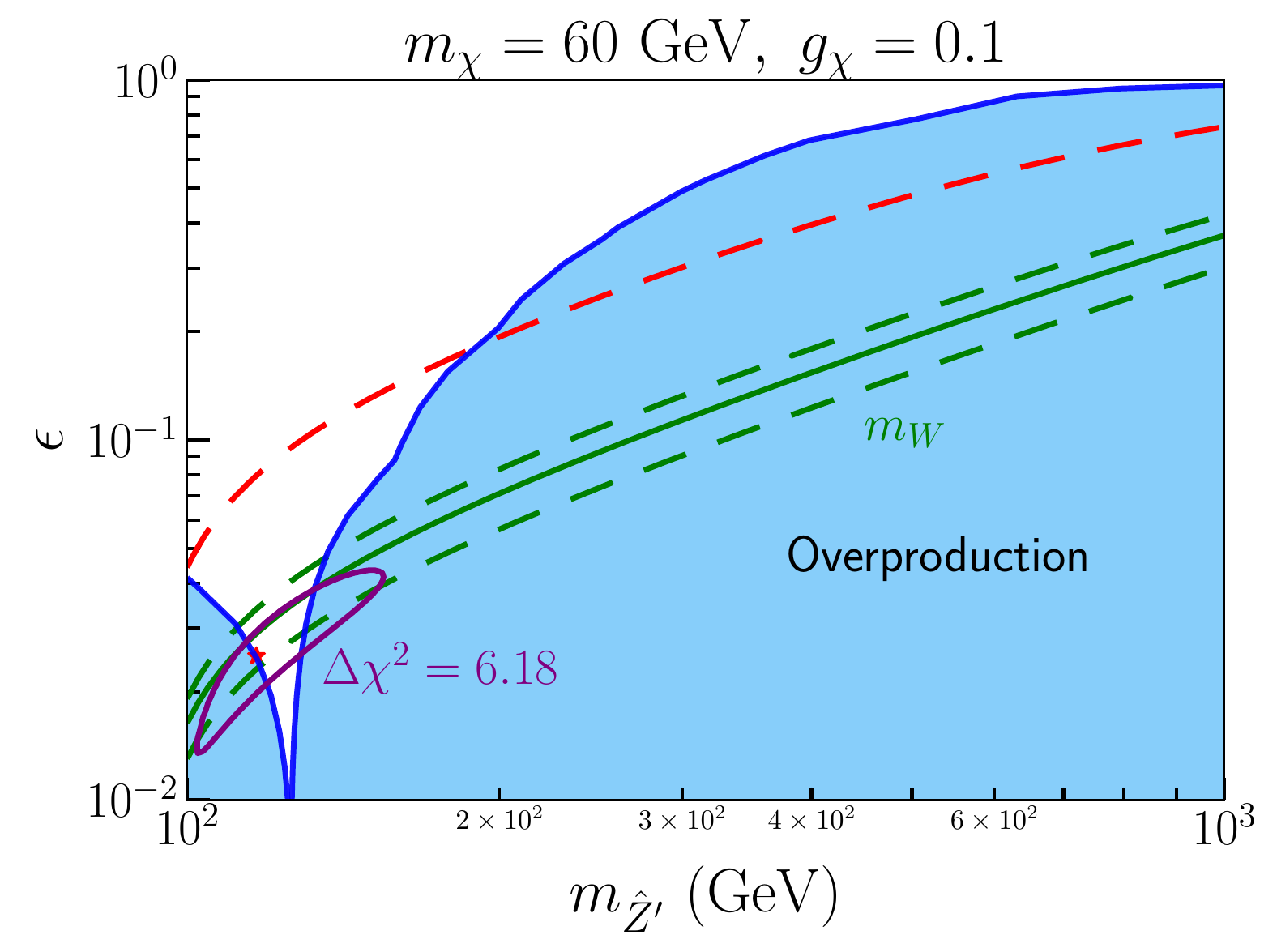}
    \caption{The lightblue area is excluded by Planck experiment~\cite{Planck:2018vyg}. The blue line gives the observed DM relic density. The red line gives the observed $W$ at tree level. The green line has taken the SM model loop corrections into consideration and gives the observed $W$ mass, with the dashed green lines correspond to the $3\sigma$ upper and lower deviation. The red star gives the best fit of electroweak oblique parameters $STU$, and the purple line corresponds to $\Delta\chi^2=6.18$ with respect to the best fit.}
    \label{fig:DPDMres}
\end{figure}

\begin{figure}[ht]
    \centering
    \includegraphics[width=0.8\textwidth]{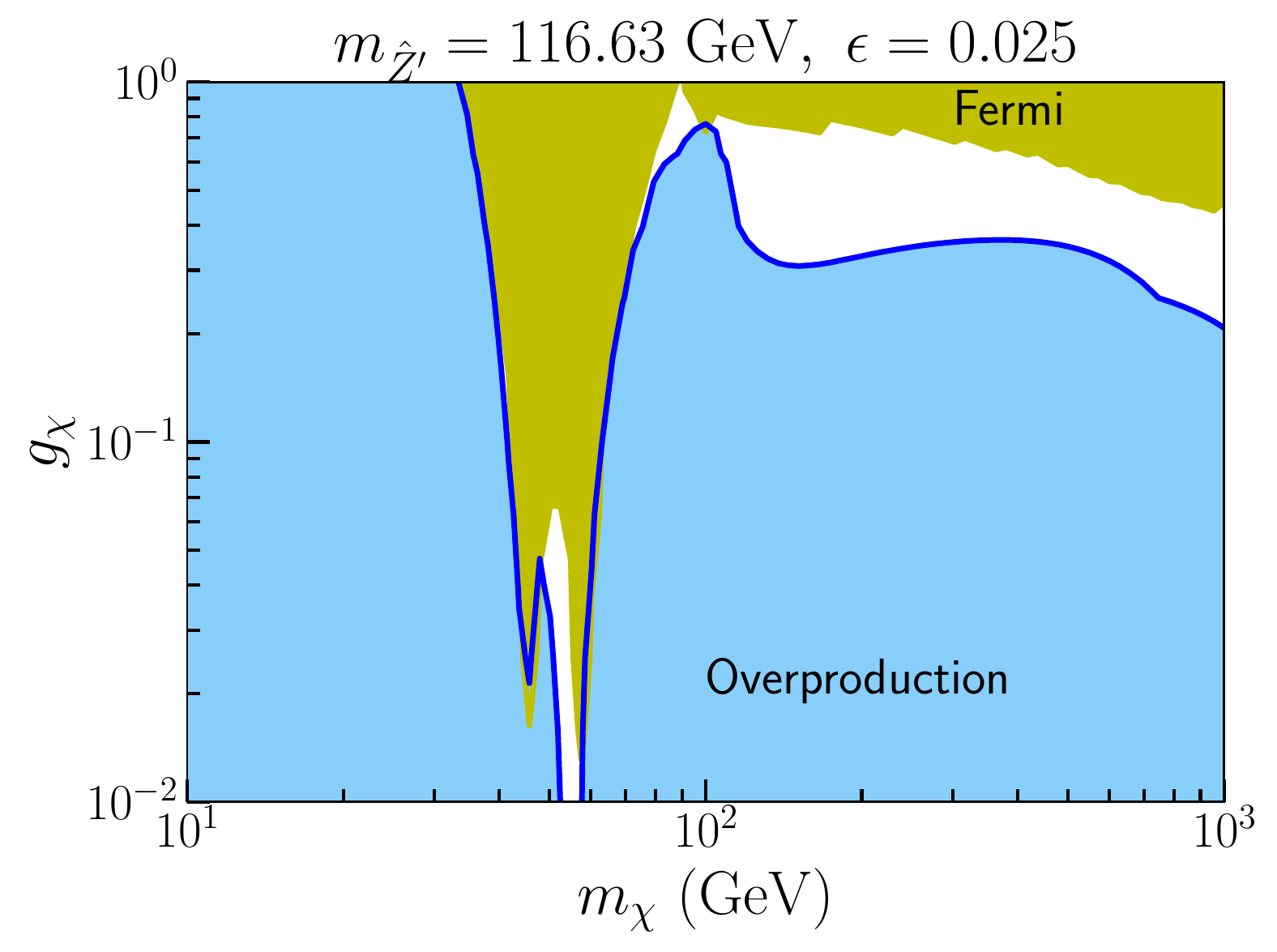}
    \caption{Constraints from DM relic density and DM indirect detection, where light blue areas are excluded by the Planck experiment~\cite{Planck:2018vyg} and yellow area is excluded by the Fermi-LAT experiment~\cite{Fermi-LAT:2016uux}. Blue lines correspond to the observed DM relic density.}
    \label{fig:DPDMindd}
\end{figure}

The DPDM model can naturally escape stringent constraint from DM direct detection due to a cancellation mechanism~\cite{Cai:2021evx,Zeng:2022llh}: the scattering between DM and SM fermions mediated by the $Z$ and $Z^{\prime}$ bosons will cancel out in the zero momentum transfer limit (i.e., the scattering amplitude is proportional to the transferred momentum). Since in $t$ channel the derivative of mediators in momentum space is proportional to the transferred momentum, models where dark sector linked to SM by the derivative of mediators will possess cancellation mechanism. In the DPDM model both spin dependent and spin independent direct detection interaction are mediated by derivative of mediators, therefore both process are suppressed in the DPDM model. In Fig.~\ref{fig:DPDMres} we have drawn the constraints from observed DM relic density, the observed $W$ mass and the electroweak oblique parameters. 
In Fig.~\ref{fig:DPDMres} the red line gives the observed $W$ boson mass solely. The green line gives the observed $W$ boson mass with SM loop corrections taken into consideration. The dashed green lines correspond to the $3\sigma$ mass deviated from the $W$ boson mass (i.e. $80.4335\pm 3\times 0.0094~\mathrm{GeV}$). The blue line saturates the observed DM relic density, while the light blue area is excluded by Planck experiment~\cite{Planck:2018vyg}. The valley of the blue line arises from the enhancement of the DM annihilation by $Z^{\prime}$ resonance. The DM relic density is calculated in settings $m_{\chi}=60~\mathrm{GeV},\ g_{\chi}=0.1$ by numerical tools: \texttt{FeynRules~2}~\cite{Alloul:2013bka}, \texttt{MadGraph}~\cite{Alwall:2014hca}, and \texttt{MadDM}~\cite{Ambrogi:2018jqj}. We use the following definition of $\chi^2$ to fit $m_{\hat{Z}^{\prime}}$ and $\epsilon$ through $STU$:
\begin{eqnarray}
    \chi^2=X Cov^{-1}X^{T},  \label{chi2} 
\end{eqnarray}
where 
\begin{eqnarray}
    X=\begin{pmatrix}
	S-0.005&T-0.04&U-0.134
    \end{pmatrix}\\
    Cov=\begin{pmatrix}
	0.097^2&\rho_{ST}\times 0.097\times 0.12&\rho_{SU}\times 0.097\times 0.087\\
	\rho_{ST}\times 0.097\times 0.12&0.12^2&\rho_{TU}\times 0.12\times 0.087\\
	\rho_{SU}\times 0.097\times 0.087&\rho_{TU}\times 0.12\times 0.087&0.087^2
    \end{pmatrix}.
\end{eqnarray}
The degree of freedom (d.o.f) of the $\chi^2$ is 3. The red star represents the best fit of $STU$: $m_{\hat{Z}^{\prime}}=116.63~\mathrm{GeV},\ \epsilon= 0.025,\ \chi^2=3.21 $. $\Delta\chi^2=6.18$ with respect to the best fit value is denoted by the purple line. From Fig.~\ref{fig:DPDMres} we see that the red star lies in the area circled by green lines, which means that the global fit of $STU$ has encoded the information of the $W$ boson mass. Also it is clear that the best fit point meets the observed DM relic density. The purple line indicates there is large area which can give explanation to the $W$ boson mass. To make the parameters fall into the purple circle $m_{\hat{Z}^{\prime}}$ should satisfy $102~\mathrm{GeV}\lesssim m_{\hat{Z}^{\prime}}\lesssim 155~\mathrm{GeV}$. To make area where $m_{\hat{Z}^{\prime}}\lesssim 114~\mathrm{GeV}$ and $m_{\hat{Z}^{\prime}}\gtrsim 132~\mathrm{GeV}$ not excluded by Planck experiment, one can change the DM mass $m_{\chi}$  and thus the annihilation resonance area will move accordingly. On the other hand, one can increase the extra gauge coupling $g_{\chi}$ or simply not introduce dark matter in this model. 
In the parameters setting adopted by Fig.~\ref{fig:DPDMres}, it is hard to find area which is not excluded by DM indirect detection. However, by switching $m_{\chi}$ and $g_{\chi}$ into free parameters and fixing $m_{\hat{Z}^{\prime}}$ and $\epsilon$, one can find large areas that escape constraints from $W$ boson mass, DM relic density and DM indirect detection at the same time. For example, fixing $m_{\hat{Z}^{\prime}}$ and $\epsilon$ to the best fit values obtained above ($m_{\hat{Z}^{\prime}}=116.63~\mathrm{GeV},\ \epsilon= 0.025$), we draw constraints on the DPDM model from DM relic density and DM indirect detection in Fig.~\ref{fig:DPDMindd}.
When the DM mass is larger than the $W$ boson mass, DM annihilate largely into $W$ bosons. When the DM mass is smaller than the $W$ boson mass, DM annihilate largely into $s$ quarks. Therefore we use these two annihilation channels to show the DM indirect detection constraints on the DPDM model. The DM annihilation cross section and DM indirect detection constraints are obtained from \texttt{MadDM} and Fermi-LAT experiment~\cite{Ambrogi:2018jqj,Fermi-LAT:2016uux}. In Fig.~\ref{fig:DPDMindd} the light blue areas and the yellow area are excluded by the Planck experiment and Fermi-LAT experiment respectively, and the blue lines correspond to the observed DM relic density.

Note that collider experiments have set strong constraints on extra vector boson~\cite{ATLAS:2017fih,ATLAS:2019erb,ParticleDataGroup:2020ssz}. For $m_{\hat{Z}^{\prime}}<209~\mathrm{GeV}$, the LEP-II experiment implies the couplings between $Z^{\prime}$ boson and the SM fermions are smaller than or of order $10^{-2}$~\cite{ParticleDataGroup:2020ssz}. The couplings between the $Z^{\prime}$ boson and the SM fermions in the DPDM model can be written as:
\begin{eqnarray}
    L_{NC,\hat{Z}^{\prime}ff}&&=\sum\limits_{f} \hat{Z}_{\mu}^{\prime}\bar{f}\gamma^{\mu}(g_{V}^{\prime}-g_{A}^{\prime}\gamma^{5})f\\
    &&=\sum\limits_{f}  (-k_1\sin \theta +k_2\cos \theta) \hat{Z}_{\mu}^{\prime}\bar{f}\gamma^{\mu}(g_{V}-g_{A}\gamma^{5})f\\
	      &&=\sum\limits_{f}  (-k_1\sin \theta +k_2\cos \theta) \hat{Z}_{\mu}^{\prime}\bar{f}\gamma^{\mu}\frac{e}{s_{w}c_{w} }(T^{3}_{f}\frac{1-\gamma^{5}}{2}-Q_{f}s_{w}^2 )f.
\end{eqnarray}
Taking $f$ to be electron we can calculate $g_{V}^{\prime}=0.0005$ and $g_{A}^{\prime}=0.0073$ for $m_{\hat{Z}^{\prime}}=116.63~\mathrm{GeV},\ \epsilon=0.025$. Therefore in the parameters setting we considered the DPDM model is safe from the LEP constraint. For hadron colliders the upper limits of couplings between the $Z^{\prime}$ boson and quarks are of order $10^{-1}$(see summary of bounds in Fig. 88.2 of Ref.~\cite{ParticleDataGroup:2020ssz}), which is relatively weaker constraints since the $Z^{\prime}$ coupling to the leptons and quarks are similar and of the same order.   

For clarity, we take $m_{\hat{Z}^{\prime}}=116.63~\mathrm{GeV},\ \epsilon=0.025,\ m_{\chi}=1000~\mathrm{GeV},\ g_{\chi}=0.21$ as a benchmark point and summarize phenomenological constraints for the benchmark point in Table~\ref{tab:bmpcons}, where the direct detection constraint on DM-xenon scattering events is calculated with the same procedure as Ref.~\cite{Zeng:2021moz}. Here for illustration we only considered the spin-independent direct detection scattering events, which is extremely small as expected by the cancellation mechanism of the DPDM model. 
\begin{table}[ht]
    \centering
    \caption{Summary of phenomenological constraints for the benchmark point $m_{\hat{Z}^{\prime}}=116.63~\mathrm{GeV},\ \epsilon=0.025,\ m_{\chi}=1000~\mathrm{GeV},\ g_{\chi}=0.21$.}
    \begin{tabular}{c c c c c c c}
   \hline
&$m_{W}$&$\chi^2$ of $STU$&$\Omega_{DM}h^2$  \\ 
   \hline
	model value&80.4136&3.21&0.1235 \\
	constraint&$80.4335\pm 0.0094~\mathrm{GeV}$~\cite{CDF:2022hxs}&d.o.f=3&0.1200$\pm$ 0.0012~\cite{Planck:2018vyg}\\
   \hline
   &$<\sigma_{ann}v>$ &$Z^{\prime}$-electron couplings&DM-Xe scattering events \\
	model value &$4.5\times 10^{-26}~\mathrm{cm}^3/\mathrm{s}$&$g_{V}^{\prime}=0.0005$ and $g_{A}^{\prime}=0.0073$&$1.2\times 10^{-9}$ \\
	constraint &$2.3\times 10^{-25}~\mathrm{cm}^3/\mathrm{s}$~\cite{Ambrogi:2018jqj,Fermi-LAT:2016uux}&$\sim O(10^{-2})$~\cite{ParticleDataGroup:2020ssz}&7.9~\cite{PandaX-4T:2021bab} \\
   \hline
    \end{tabular}
    \label{tab:bmpcons}
\end{table}
\subsection{U(1) model}
\label{sub:u_1_model}
In the U(1) model there is a gauge boson of an extra $\mathrm{U}(1)_\mathrm{X}$ gauge symmetry which connects to the gauge boson of SM $\mathrm{U}(1)_\mathrm{Y}$ symmetry through kinetic mixing. In this section we will adopt the same model setting as Ref.~\cite{Lao:2020inc}. Then the kinetic mixing terms can be written as:
\begin{eqnarray}
    \mathcal{L}_{\mathrm{K}}=-\frac{1}{4}B^{\mu\nu}B_{\mu\nu}-\frac{1}{4}X^{\mu\nu}X_{\mu\nu}-\frac{\epsilon}{2}B^{\mu\nu}X_{\mu\nu},
\end{eqnarray}
where $B_{\mu}$ and $X_{\mu}$ are the gauge fields of $\mathrm{U}(1)_\mathrm{Y}$ and $\mathrm{U}(1)_\mathrm{X}$ gauge symmetry. After Higgs filed getting its vev there will be mass mixing term between $B_{\mu}$ and $W^{3}_{\mu}$, and then the matrix of $(W_{\mu}^{3},\ B_{\mu},\ X_{\mu}  )$ can be denoted as:
\begin{eqnarray}
   &&\frac{1}{2}\begin{pmatrix}
 W^{3\mu}&B^{\mu}&X^{\mu}      
   \end{pmatrix}
    \begin{pmatrix}
	g^2v^2/4&-gg^{\prime}v^2/4&0\\
	-gg^{\prime}v^2/4&g^{\prime 2}v^2/4&0\\
	0&0&g_{x}^2v_{s}^2
    \end{pmatrix}
   \begin{pmatrix}
 W_{\mu}^{3}\\
 B_{\mu}\\
 X_{\mu}      
   \end{pmatrix}\nonumber\\
    =&&\frac{1}{2}\begin{pmatrix}
 W^{3\mu}&B^{\mu}&X^{\mu}      
   \end{pmatrix}K^{-1T}OO^TK^{T}
    \begin{pmatrix}
	g^2v^2/4&-gg^{\prime}v^2/4&0\\
	-gg^{\prime}v^2/4&g^{\prime 2}v^2/4&0\\
	0&0&g_{x}^2v_{s}^2
    \end{pmatrix}KOO^T K^{-1}
   \begin{pmatrix}
 W_{\mu}^{3}\\
 B_{\mu}\\
 X_{\mu}      
   \end{pmatrix}\label{massmatrix}\\
    =&&\frac{1}{2}\begin{pmatrix}
 A^{\mu}&\hat{Z}^{\mu}&\hat{Z}^{\prime\mu}      
   \end{pmatrix}
    \begin{pmatrix}
	0&0&0\\
	0&m_{\hat{Z}}^2&0\\
	0&0&m_{\hat{Z}^{\prime}}^2
    \end{pmatrix}
   \begin{pmatrix}
 A_{\mu}\\
 \hat{Z}_{\mu}\\
 \hat{Z}^{\prime}_{\mu}      
   \end{pmatrix},\nonumber
\end{eqnarray}
where $g_{x}$ is gauge coupling of the $\mathrm{U}(1)_\mathrm{X}$ gauge symmetry and $v_{s}$ is the vev of an dark scalar which gives mass to $X_{\mu}$. In Eq.~\eqref{massmatrix} we have used $K$ to normalize the kinetic terms of $B_{\mu}$ and $X_{\mu}$ and used $O$ to diagonalize the mass matrix and transform the fields to their mass eigenstates. The masses of the two massive vector boson $\hat{Z}$ and $\hat{Z}^{\prime}$ are:
\begin{eqnarray}
    m_{\hat{Z},\hat{Z}^{\prime}}^2&&=
    \frac{1}{8} (
    g^2 v^2+g^{\prime 2} k_1^2 v^2+g^{\prime 2} k_2^2 v^2+4 g_x^2 k_1^2 v_s^2+4 g_x^2 k_2^2 v_s^2\\
				  &&\pm\sqrt{\left(g^2 v^2+\left(k_1^2+k_2^2\right) \left(g^{\prime 2} v^2+4 g_x^2 v_s^2\right)\right)^2-16 g_x^2 v^2 v_s^2 \left(g^2 \left(k_1^2+k_2^2\right)+4 g^{\prime 2} k_1^2 k_2^2\right)}),\nonumber
\end{eqnarray}
where we denote the masses and mass eigenstates with hat in order to keep consistent with the DPDM model.
Note that the kinetic mixing between $B_{\mu}$ and $X_{\mu}$ will not change the form of the electric charge $e$. The definition of electric can be extracted from couplings between the photon and the Higgs doublet. Which in this model will be:
\begin{eqnarray}
    e=g[KO]_{11}=g^{\prime}[KO]_{21}=\frac{2g^{\prime}k_2}{\sqrt{1+\frac{4g^{\prime 2}k_2^2}{g^2}+\frac{k_2^2}{k_1^2}} }=\frac{gg^{\prime}}{\sqrt{ g^2+g^{\prime 2}}},
\end{eqnarray}
where $[KO]_{ij}$ represents the element which lies in the $i^{th}$ row and the $j^{th}$ column of matrix $KO$. 
The neutral-current coupling between $Z$ boson and SM fermions in the U(1) model can be written as:
\begin{eqnarray}
    L_{NC,\hat{Z}ff}&&=\sum\limits_{f}\hat{Z}_{\mu}\bar{f}\gamma^{\mu}(g_{V}-g_{A}\gamma^{5})f,\\
    \mathrm{with}\ g_{V}&&=g_{A}+g^{\prime}[KO]_{22}Q_{f},\ g_{A}=\frac{T_{f}^3}{2}(-g^{\prime}[KO]_{22}+g[KO]_{12}).
\end{eqnarray}
From the above expression we can read $S,T,U$ as:
\begin{eqnarray}
    \alpha T=\frac{2\hat{s}_{w}\hat{c}_{w}(-g^{\prime}[KO]_{22}+g[KO]_{12})}{e}-2\\
    \alpha S=\frac{-4g^{\prime}[KO]_{22}(\hat{c}_{w}^2-\hat{s}_{w}^2)}{-g^{\prime}[KO]_{22}+g[KO]_{12}}-4\hat{s}_{w}^2(\hat{c}_{w}^2-\hat{s}_{w}^2)+4\hat{c}_{w}^2\hat{s}_{w}^2\alpha T\\
    \alpha U=8\hat{s}_{w}^2(\frac{\hat{s}_{w}}{s_{w}}-1+\frac{\alpha S}{4(\hat{c}_{w}^2-\hat{s}_{w}^2)}-\frac{\hat{c}_{w}^2\alpha T}{2(\hat{c}_{w}^2-\hat{s}_{w}^2)}).
\end{eqnarray}

\begin{figure}[ht]
    \centering
    \includegraphics[width=0.8\textwidth]{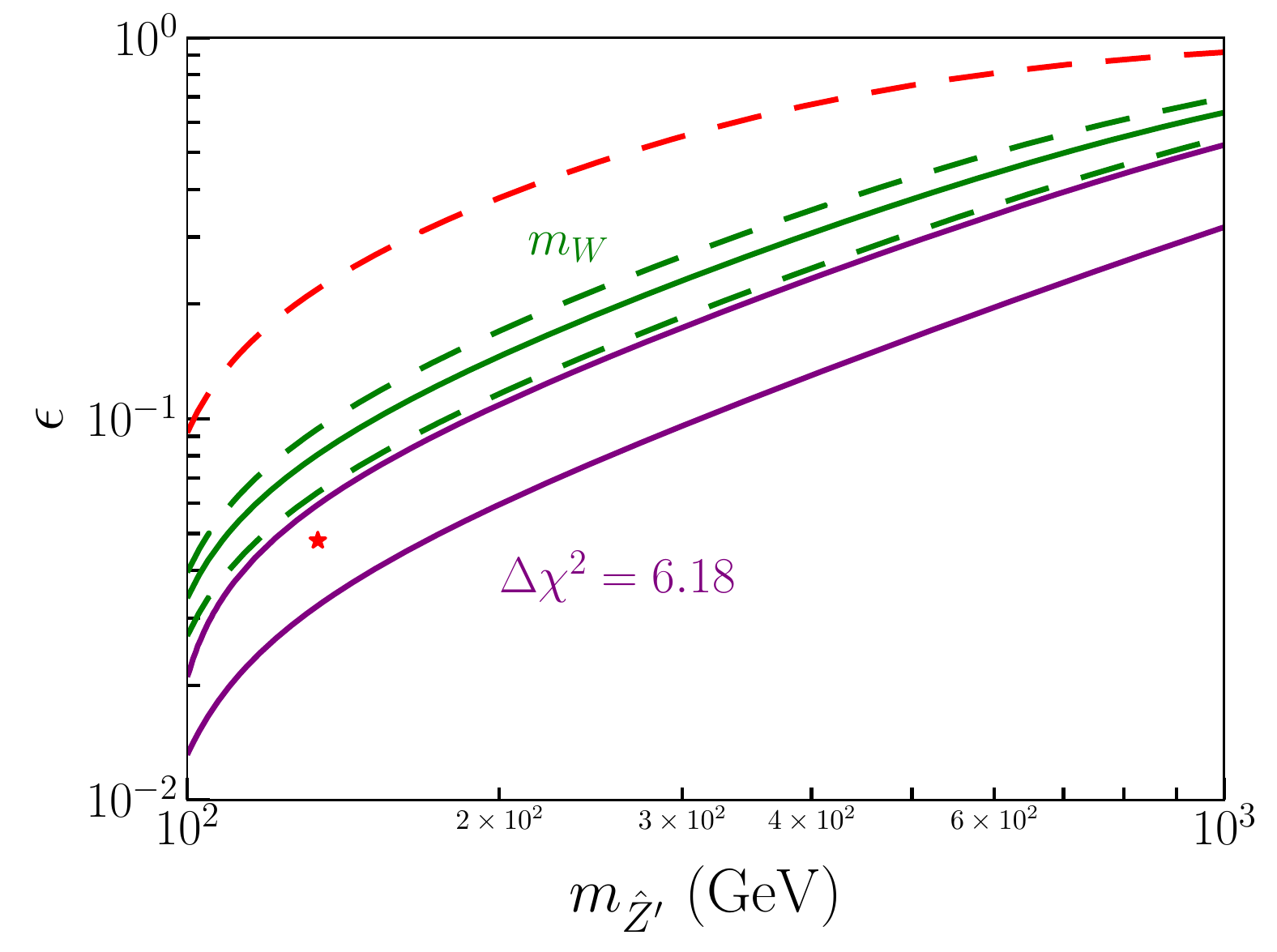}
    \caption{The red line gives the observed $W$ boson mass at tree level. The green line has taken the SM model loop corrections into consideration and gives the observed $W$ boson mass, with the dashed green lines correspond to the $3\sigma$ upper and lower deviation. The red star gives the best fit of the electroweak oblique parameters $STU$, and the purple lines correspond to $\Delta\chi^2=6.18$.    }
    \label{fig:u1res}
\end{figure}

Now we can give a line which predicts the observed $W$ boson mass in this model in Fig.~\ref{fig:u1res}.
In Fig.~\ref{fig:u1res} we also use red dashed line to show that the U(1) model can solely give the observed mass of the $W$ boson. The green line takes the SM loop corrections into consideration and gives the observed $W$ boson mass, with the dashed green lines correspond to the $3\sigma$ upper and lower deviation. We use the same definition of $\chi^2$ as Eq.~\eqref{chi2}. The red star being the best fit of electroweak oblique parameters $STU$ : $m_{\hat{Z}^{\prime}}=133.65~\mathrm{GeV},\ \epsilon=0.048,\ \chi^2=24.94$, with the purple lines corresponding to $\Delta\chi^2=6.18$.  From Fig.~\ref{fig:u1res} we see that the electroweak oblique parameters results do not fall into the area circled by the green lines which represents the direct calculation of $m_{W}$. This means that the parameters space which gives the observed $W$ boson mass can not give corresponding electroweak couplings that fit the electroweak oblique parameters nicely. Our results shows that the best fit of the U(1) model can only give about $27~\mathrm{MeV}$ extra mass to the SM $W$ boson mass. 
See Appendix~\ref{sec:appendix} for comparison of electroweak oblique parameters between the DPDM model and the U(1) model. 
\section{\label{sec:con}Conclusion}
In this work we have explored the possibility of altering the $W$ boson mass at tree level through mixing between an extra gauge boson and the $Z$ boson. We first gave general discussion of the effects from mixing extra vector boson with the $Z$ boson, then explored two realistic models: the DPDM model and the U(1) model. In the DPDM model the extra gauge boson mixes with the $Z$ boson through the kinetic mixing between the extra boson and the $Z$ boson, while in the U(1) model the extra gauge boson mixes with the $Z$ boson through the kinetic mixing between the extra boson and the $B$ boson. Apart from giving the $W$ boson mass, we also discussed the electroweak oblique parameters constraints for both models, and explored DM relic density and DM indirect detection constraints for the DPDM model. We find that in both model the best fit value for the extra vector boson mass is around $120~\mathrm{GeV}$. While the best fit of the U(1) model can only contribute $27~\mathrm{MeV}$ extra mass to the SM $W$ boson mass, the best fit of the DPDM model can give the observed $W$ boson mass as well as the observed DM relic density. The DPDM model can also escape stringent DM direct detection and the best fit of the DPDM can saturate the constraints from DM indirect detection and rough estimation of collider bounds. Detailed collider search for the DPDM model seemed interesting and is leaved for future works.

\begin{acknowledgments}
    This work is supported in part by the National Natural Science Foundation of China (NSFC) under Grant Nos. 11875327 and 11905300, the China Postdoctoral Science Foundation under Grant
No. 2018M643282, the Fundamental Research Funds for the Central Universities, and the Sun Yat-Sen University Science Foundation.
\end{acknowledgments}
\section*{Note added}
During the finalizing of this manuscript, we noticed that Ref.~\cite{Zhang:2022nnh} appeared on arxiv. Ref.~\cite{Zhang:2022nnh} discusses explanation of the $W$ boson mass with U(1) dark matter model as well as several phenomenology constraints on DM. Our work discusses models with an extra gauge boson which can explain the $W$ boson mass. Apart from the DPDM model, we also discussed the U(1) model, but in different scenarios.
\appendix
\section{\label{sec:appendix}Comparison of $STU$ between the DPDM model and the U(1) model }
To see why the $STU$ behave differently in the DPDM model and the U(1) model analytically, we compare the expression and structure of the $STU$ between these two models in this section.

In Sec.~\ref{sub:u_1_model} we gave the general expression of $STU$ of the U(1) model, here we will show the exact expression of these variables. To do this we take $K$ in Eq.~\eqref{massmatrix} as:
\begin{eqnarray}
    K=\begin{pmatrix}
	1&0&0\\
	0&1&-t_{\epsilon}\\
	0&0&1/c_{\epsilon}
    \end{pmatrix},
\end{eqnarray}
where $c_{\epsilon}=\sqrt{1-\epsilon^2} $ and $t_{\epsilon}=\epsilon/c_{\epsilon}$. Then $O$ in Eq.~\eqref{massmatrix} can be denoted as:
\begin{eqnarray}
    O=O_{w}O_{\xi}=
	\begin{pmatrix}
	    s_{w}&-c_{w}&0\\
	    c_{w}&s_{w}&0\\
	    0&0&1
	\end{pmatrix}
	\begin{pmatrix}
	    1&&\\
	     &c_{\xi}&s_{\xi}\\
	     &-s_{\xi}&c_{\xi}
	\end{pmatrix},
\end{eqnarray}
where $s_{w},\ c_{w},\ s_{\xi} $ and $ c_{\xi}$ are shorthand notations for $\sin \theta_{w},\ \cos \theta_{w},\ \sin \xi $ and $ \cos \xi$ respectively.
Substituting $K$ and $O$ into the expression of the $STU$ of the U(1) model and after simplification we arrive:
 \begin{eqnarray}
     \alpha T&&=-2c_{\xi}\frac{s_{w}c_{w}}{\hat{s}_{w}\hat{c}_{w}}-2\label{u_1aT}\\
	      \alpha S&&
    =4(\hat{c}_{w}^2-\hat{s}_{w}^2)(s_{w}^2-\hat{s}_{w}^2+\frac{s_{w}^2c_{w}^2-\hat{s}_{w}^2\hat{c}_{w}^2}{s_{w}^2})+4\hat{c}_{w}^2\hat{s}_{w}^2\alpha T\\
    \alpha U&&=8\hat{s}_{w}^2(\frac{\hat{s}_{w}}{s_{w}}-1+\frac{\alpha S}{4(\hat{c}_{w}^2-\hat{s}_{w}^2)}-\frac{\hat{c}_{w}^2\alpha T}{2(\hat{c}_{w}^2-\hat{s}_{w}^2)})\\
	    &&=8\hat{s}_{w}^2(\frac{\hat{s}_{w}}{s_{w}}-1+(s_{w}^2-\hat{s}_{w}^2+\frac{s_{w}^2c_{w}^2-\hat{s}_{w}^2\hat{c}_{w}^2}{s_{w}^2}))-4\hat{s}_{w}^2\hat{c}_{w}^2\alpha T.
 \end{eqnarray}
For comparison we write down the $STU$ of the DPDM model:
\begin{eqnarray}
    \alpha T&&=2(\frac{\hat{s}_{w}\hat{c}_{w}}{s_{w}c_{w}}(-k_2\sin \theta-k_1\cos \theta)-1)\\
    \alpha S&&=4\hat{c}_{w}^2\hat{s}_{w}^2\alpha T+4(\hat{c}_{w}^2-\hat{s}_{w}^2)(s_{w}^2-\hat{s}_{w}^2)\\ 
    \alpha U&&=8\hat{s}_{w}^2(\frac{\hat{s}_{w}}{s_{w}}-1+\frac{\alpha S}{4(\hat{c}_{w}^2-\hat{s}_{w}^2)}-\frac{\hat{c}_{w}^2\alpha T}{2(\hat{c}_{w}^2-\hat{s}_{w}^2)})\\
    &&=8\hat{s}_{w}^2(\frac{\hat{s}_{w}}{s_{w}}-1+(s_{w}^2-\hat{s}_{w}^2))-4\hat{s}_{w}^2\hat{c}_{w}^2\alpha T.
\end{eqnarray}
We see that the $STU$ expression of these two model are similar. Knowing that $\Delta s=\hat{s}_{w}-s_{w}$ is small, we can expand $S$ and $U$ to $O(\Delta s)$ and neglect the higher order terms. Then for the U(1) model
\begin{eqnarray}
     \alpha T&&=-2c_{\xi}\frac{s_{w}c_{w}}{\hat{s}_{w}\hat{c}_{w}}-2\\
	      \alpha S&&
    =-8\hat{c}_{w}^2(\hat{c}_{w}^2-\hat{s}_{w}^2)\frac{\Delta s}{\hat{s}_{w}}+4\hat{c}_{w}^2\hat{s}_{w}^2\alpha T\\
    \alpha U
		      &&=-8\hat{s}_{w}^2(\hat{c}_{w}^2-\hat{s}_{w}^2)\frac{\Delta s}{\hat{s}_{w}}-4\hat{s}_{w}^2\hat{c}_{w}^2\alpha T,
\end{eqnarray}
while for the DPDM model
\begin{eqnarray}
    \alpha T&&=2(\frac{\hat{s}_{w}\hat{c}_{w}}{s_{w}c_{w}}(-k_2\sin \theta-k_1\cos \theta)-1)\\
    \alpha S&&=4\hat{c}_{w}^2\hat{s}_{w}^2\alpha T-8\hat{s}_{w}(\hat{c}_{w}^2-\hat{s}_{w}^2)\Delta s\\ 
    \alpha U&&=8\hat{s}_{w}(\hat{c}_{w}^2-\hat{s}_{w}^2)\Delta s-4\hat{s}_{w}^2\hat{c}_{w}^2\alpha T.
\end{eqnarray}
It is interesting to see that the first order expression of $S$ and $U$ in the DPDM model are the opposite of each other. Actually $\hat{s}_{w}$ and $\hat{c}_{w}$ can be calculated from Eq.~\eqref{swcwrel}, therefore we can further simplify the expression of the $STU$ as
\begin{eqnarray}
      T&&=-\frac{2c_{\xi}s_{w}c_{w}}{\alpha\hat{s}_{w}\hat{c}_{w}}-\frac{2}{\alpha},\ S=-866.17\Delta s+0.716 T,\ U=-263.76\Delta s-0.716 T
\end{eqnarray}
and 
\begin{eqnarray}
     T&&=(\frac{2\hat{s}_{w}\hat{c}_{w}(-k_2\sin \theta-k_1\cos \theta)}{s_{w}c_{w}\alpha}-\frac{2}{\alpha})\\
    S&&=0.716 T-263.76\Delta s,\ U=263.76\Delta s-0.716 T.
\end{eqnarray}
$\Delta s$ represents the deviation of $s_{w}$ from its SM value $\hat{s}_{w}$. Taking $s_{w}=0.48269$, which gives the desired $g$ (thus the desired $m_{W}$), we can obtain $\Delta s=0.00046$. Therefore the $S$ and $U$ of the U(1) model and the DPDM model can be written as $S=0.716T-0.398,\ U=-0.716T-0.121$ and $S=0.716T-0.121,\ U=-0.716T+0.121$ respectively. From Eq.~\eqref{u_1aT} we see the $T$ in the U(1) model will always be negative, which means in the U(1) model the $S$ will be smaller than $-0.398$ when the $W$ boson mass is satisfied. In this case the $S$ deviates largely from its central value $0.005$, therefore to fulfill the electroweak oblique parameters fit the region of large $W$ boson mass enhancement will be disfavored (as we see in Fig.~\ref{fig:u1res} that the best fit of the U(1) model can only provide about $27~\mathrm{MeV}$ extra mass to the SM $W$ boson mass). As a comparison, the $S$ of the DPDM model does not deviate too much from its central value. Also the $U$ of the DPDM model has positive value, and thus it is easier to approach it large central value $0.134$. Therefore it is reasonable to see the DPDM model to fulfill the oblique electroweak parameters fit and the $W$ boson mass simultaneously. 

To further understand why the contours of these two model behave differently, we further write down the $T$ of the U(1) model as:
\begin{eqnarray}
    T=\frac{2}{\alpha} \sqrt{\frac{t_{\epsilon}^2s_{w}^2}{(x^2-1)^2+t_{\epsilon}^2s_{w}^2}}x -\frac{2}{\alpha},
\end{eqnarray}
where $x=\frac{s_{w}c_{w}}{\hat{s}_{w}\hat{c}_{w}}$. Noticing $(x^2-1)^2$ is much smaller than $t_{\epsilon}^2s_{w}^2$ in the region we studied, we can expand $T$ to $O(\frac{(x^2-1)^2}{t_{\epsilon}^2s_{w}^2})$ as:
\begin{eqnarray}
    T=\frac{2}{\alpha}(1-\frac{1}{2}\frac{(x^2-1)^2}{t_{\epsilon}^2s_{w}^2})x-\frac{2}{\alpha}.
\end{eqnarray}
From Fig.~\ref{fig:u1res} we see that the upper purple line is similar and close to the lower dashed green line. All points in the lower dashed green line give the same $W$ boson mass, which means the same $g$ and the same $s_{w}$. Therefore we expect points in the purple line have similar $s_{w}$. If along the purple line the $T$ are almost the same, then $S$ and $U$ will also have little changes along the purple line since they are determined by $s_{w}$ and $T$. Then it is reasonable to see the $\chi^2$ contour (the purple line) and the $m_{W}$ contour (the dashed green line) behave similarly.   
Taking $s_{w}=0.48286$ (the lower dashed green line value) as an approximation and $\epsilon=0.2$ as a benchmark, we obtain $\frac{1}{2}\frac{(x^2-1)^2}{t_{\epsilon}^2s_{w}^2}=3.64\times10^{-5}$ and $x=1-4.21\times10^{-4}$. We see $T$ can be rewritten as:
\begin{eqnarray}
    T&&=\frac{2}{\alpha}(-\frac{1}{2}\frac{(x^2-1)^2}{t_{\epsilon}^2s_{w}^2}-(1-x)+\frac{1}{2}\frac{(x^2-1)^2}{t_{\epsilon}^2s_{w}^2}(1-x))\\
    &&\approx \frac{2}{\alpha}(-\frac{1}{2}\frac{(x^2-1)^2}{t_{\epsilon}^2s_{w}^2}-(1-x)).
\end{eqnarray}
$\frac{1}{2}\frac{(x^2-1)^2}{t_{\epsilon}^2s_{w}^2}$ decreases as $\epsilon$ increases, and when $\epsilon=0.5$ it decreases to $4.55\times10^{-6}$. We see the $T$ do change little when $s_{w}$ is fixed and $\epsilon$ increases. When $\epsilon$ decreases, $\frac{1}{2}\frac{(x^2-1)^2}{t_{\epsilon}^2s_{w}^2}$ can be comparable or larger than $4.21\times 10^{-4}$, therefore to keep $\chi^2$ unchanged $s_{w}$ should change accordingly. From Fig.~\ref{fig:u1res} it is obvious to see $s_{w}$ in the lower left part of the upper purple line do increases. For the DPDM model, $T$ can be written as:
\begin{eqnarray}
    T=\frac{2}{\alpha}\frac{1}{\sqrt{1+\frac{(x^2-1)^2}{t_{\epsilon}^2} }}x-\frac{2}{\alpha}.
\end{eqnarray}
The expression of $T$ in the DPDM model is similar to that of the U(1) model, however the $\Delta \chi^2$ of the DPDM model is different from that of the U(1) model, which means in the case of the DPDM model the minor change of the $T$ will affect the $\chi^2$ value distinctively. To see this clearly we will write down the expression of $\chi^2$ explicitly:
\begin{eqnarray}
    \chi^2&&=1494.83 S^2+S (-3620.67 T-2228.11 U+428.445)+2500.26 T^2\\
	  &&+T (3445.65 U-643.634)+1415.9 U^2-506.147 U+45.7134.
\end{eqnarray}
We see $\chi^2=45.7134$ corresponds to the SM case. For the U(1) model the $\chi^2$ can be approximated as:
\begin{eqnarray}
    \chi^2&&=7.10963\times 10^8 \Delta s^2+\Delta s (-53071 T-237605)\\
    &&+75.2286 T^2+25.5335 T+45.7134.
\end{eqnarray}
While for the DPDM model the $\chi^2$ can be approximated as:
\begin{eqnarray}
    \chi^2&&=3.57505\times 10^8 \Delta s^2+\Delta s (-77147.6 T-246508)\\
	&&+75.2286 T^2+25.5335 T+45.7134.
\end{eqnarray}
We see the $\chi^2$ in the DPDM model depends more on $T$ than that of the U(1) model. Also the $\chi^2$ value of the contour in the DPDM model is one third of that in the U(1) model. Therefore the $\chi^2$ in the DPDM model will vary more drastically as $T$ varies than that in the U(1) model, resulting in a different contour shape.   
\bibliography{ref,INSPIRE-CiteAll}
\end{document}